# Forecasting Exchange Rates Using Time Series Analysis: The sample of the currency of Kazakhstan

Daniya Tlegenova
118425, Singapore
daniya.tlegenova@gmail.com

**ABSTRACT** - This paper models yearly exchange rates between USD/KZT, EUR/KZT and SGD/KZT, and compares the actual data with developed forecasts using time series analysis over the period from 2006 to 2014. The official yearly data of National Bank of the Republic of Kazakhstan is used for present study. The main goal of this paper is to apply the ARIMA model for forecasting of yearly exchange rates of USD/KZT, EUR/KZT and SGD/KZT. The accuracy of the forecast is compared with Mean Absolute Error (MAE), Mean Absolute Percentage Error (MAPE) and Root Mean Squared Error (RMSE).

**INTRODUCTION**

Exchange rate is the currency rate of one country expressed in terms of the currency of another country [1]. In the modern world, exchange rates of the most successful countries are tend to be floating. This system is set by the foreign exchange market over supply and demand for that particular currency in relation to the other currencies [2]. In addition, the exchange rate is guided by significant impact of the activities of central banks and other financial institutions.

There is a tendency that Kazakh tenge (KZT) depreciates against the major world currencies such as the United States of America dollar (USD) and euro (EUR). In particular, the National Bank of Kazakhstan made several changes to its monetary policy on September 2, 2013. Daily exchange rate was fixed by the National Bank of Kazakhstan in relation with a weighted basket of foreign currencies which include USD (70 per cent), EUR (20 per cent) and Russian ruble (10 per cent) [3]. In addition to this, in 2013 the National Bank of Kazakhstan used a substantial foreign exchange reserves to maintain the exchange rate against the background of the devaluation expectations and weakening currencies of developing countries. As a result, the real effective exchange rate proved to be overvalued, and in February, 2014 the regulator announced the failure to maintain the old exchange rate of tenge and set a new corridor of fluctuations of 185 (±3) tenge per one USD, which corresponds to one-time currency weakening by approximately 20 per cent. Moreover, in mid-July, 2015, the National Bank of Kazakhstan confirmed a new expansion of the pegged corridor of the Kazakh tenge exchange rate from 170-188 tenge per one USD [4]. Subsequently, on August 20, 2015 the country had switched to a free float and decided to pursue an inflation-targeting monetary policy. The government announced that Kazakhstan's shift to a free float "will create the necessary conditions for a recovery of economic growth, increased lending and investment activity, creation of new jobs and a decrease in the inflation rate to between 3 per cent and 4 per cent in the medium term," [5]. As a result, Kazakhstan's currency had lost more than 20 per cent of its value after the government allowed the currency to float freely [6].

Amid of a sharp drop of the exchange rate of KZT, there might be a possibility of loss to the population's savings, and probably to the country's economy as a whole. This problem may raise a question as: how can one predict the currency exchange rates' future fluctuations? In this regard, we refer to the methods of forecasting the dynamics of exchange rates.

People in Kazakhstan are likely to keep their savings in their national currency, and currency risks may arise in terms of devaluation of tenge. This economic process is defined as the process of losing money or shortfall in income due to the uncertainty of changes in exchange rates compared with the forecast values for a certain period of time [4]. Thus, reducing the risk of currency exchange rate fluctuations is to decrease uncertainty by analyzing exchange rates. Moreover, forecasts in exchange rates are essential for all aspects of the international economic relations.

It seem to be very difficult to analyze how the foreign exchange rate changes, and probably even harder to forecast them. There are lot of works done on time series based prediction modelling of foreign currency rates in literature. Many authors [7]-[11] created and tested the Autoregressive Integrated Moving Average (ARIMA) model to forecast exchange rates. Monthly or daily exchange rates were used as the variable output in these reports. These studies outlined that the ARIMA model is comparatively accurate model to forecast the exchange rate. Akincilar et al. [7] studied the exchange rate forecasting of US dollar, euro and Great Britain pound with respect to the Turkish lira. Several methods were performed for forecasting and then compared with the ARIMA model. The performance of the models was estimated via mean absolute percentage error (MAPE), root mean square errors (RMSE) and mean square error (MAE). Weisang et al. [8] further developed a detailed ARIMA modelling in the form of a case study using macroeconomic indicators to model the USD/EUR exchange rate. They developed a linear relationship for the monthly USD/EUR exchange rate over the period from January 1994 to October 2007.

Besides this, Box-Jenkins approach [12] is applied successfully in many areas such as tourism demand [13]-[17], energy [18], [19] and many others. Box and Jenkins' ARIMA technique [12] has been extensively used as a standard for time series forecasting and for evaluation of the new modelling approaches in the last twenty five years, and probably it dominates the time series forecasting. Box-Jenkins [12] paid special attention to the selecting the model and its evaluation. The methodology for constructing ARIMA model for the investigated time series includes the following main steps [12]:
1. Identification of test patterns;
2. Estimation of the model parameters and identifying the adequacy of the model;
3. The use of models to predict.

Forecasting exchange rate appears to be a very attracting topic in international finance. Objective of this paper is to apply ARIMA technique for forecasting currency exchange rates of KZT against three other currencies such as USD, EUR, and Singapore dollar (SGD) using their historical exchange rates. This prediction of the exchange rates was conducted with the help of MATLAB software.

The paper is structured as follows. Section 2 describes the Autoregressive Integrated Moving Average (ARIMA) technique. Section 3 describes the forecasting exercise and discusses the results. Section 4 summarizes and concludes.

**METHODOLOGY**

The exchange rate series USD/KZT, EUR/KZT, and SGD/KZT for the period from January 01, 2006 to December 31, 2014 [4] is exhibited in Figure 1. The historical currency exchange rates from January 2006 to December 2014 provided by the National Bank of Kazakhstan [4] were used. It is quite evident from the graph that there is an upward trend for all three currencies with respect to KZT throughout eight years.

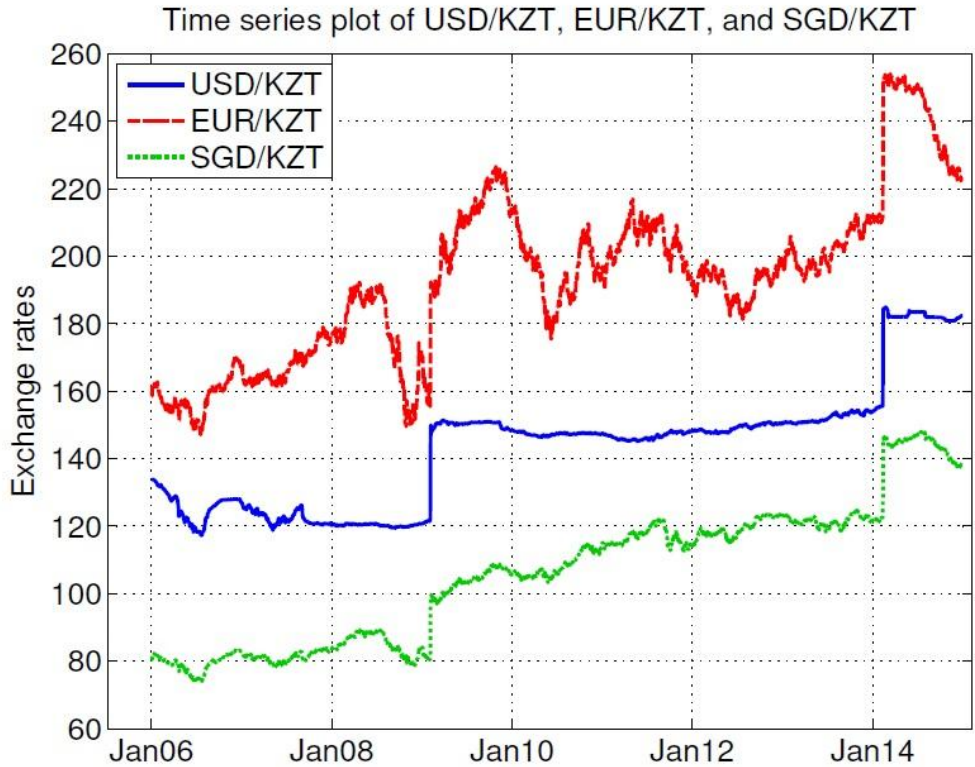

**Figure 1. Time series plot for the period from Jan 2006 to Dec 2014**

We note that the most textbooks on time series analysis [20]-[23] give the accurate and in-depth information on the methodology of ARIMA estimation and model selection. ARIMA model in our study was performed in a following way. First, it was decomposed into two parts, namely, integrated component and the ARMA model. Next, the second component (ARMA) was further decomposed into two as autoregressive (AR) and moving average (MA). The AR component related the current values of the given time series to its previous values with respect to the time. The MA component correlated the random values at current time to the previous time, by presenting the duration of its influence. The ARMA model as a combination of autoregressive and moving average models is represented in (1) [24]:

$$y_t = \beta_0 + \beta_1 y_{t-1} + \ldots + \beta_p y_{t-p} - \alpha_1 u_{t-1} - \alpha_2 u_{t-2} - \ldots - \alpha_q u_{t-q} + u_t, \qquad (1)$$

where p and q are integers greater than or equal to zero.

The three main steps of this model, such as Identification, Estimation, and Model checking are elaborated as follows. Firstly, the stationarity of the time series is established. Next, the conditional mean model for the given data is identified [24]. For AR model the sample autocorrelation function (ACF) tails off, whereas the partial autocorrelation function (PACF) cuts off after q number of lags. As opposed, for MA process the sample ACF cuts off after q lags, but the PACF tails off. In the case if both ACF and PACF tail off, proceed with the ARMA model [24]. Secondly, the model parameters are estimated by utilizing the maximum likelihood method [25]. Thirdly, the model checking is performed by diagnostics of randomness of the residuals. The residuals are required to be uncorrelated and normally distributed [26]. Finally, one can perform forecasting with the chosen model over future finite time space.

## RESULTS AND DISCUSSION

The ARIMA model is performed for forecasting of exchange rates for USD/KZT, EUR/KZT and SGD/KZT. The time period was taken from January, 2006 to December, 2014. The data was taken from the depository of the National Bank of Kazakhstan Official internet resource [4]. The mean absolute error (MAE), mean absolute percentage error (MAPE) and root mean squared error (RMSE) are selected to be the forecasting accuracy measures.

Forecasting of three foreign currency exchange rates with respect to Kazakh tenge is represented as follows. Figures 2-4 show that the time series data was not stationary. Figure 2 represents the actual and the forecast values of the USD/KZT exchange rates. As can be seen, there is a tendency of increase for the period of eight years from 2006 to 2014. For example, if the exchange rate for the year 2006 was 126.1 KZT for one USD, this value grew up to 182.19 KZT in 2014. As a result, the total increase over eight years period was more than 44 per cent. Considering this, the forecast model shows the similar trend as the actual exchange rates.

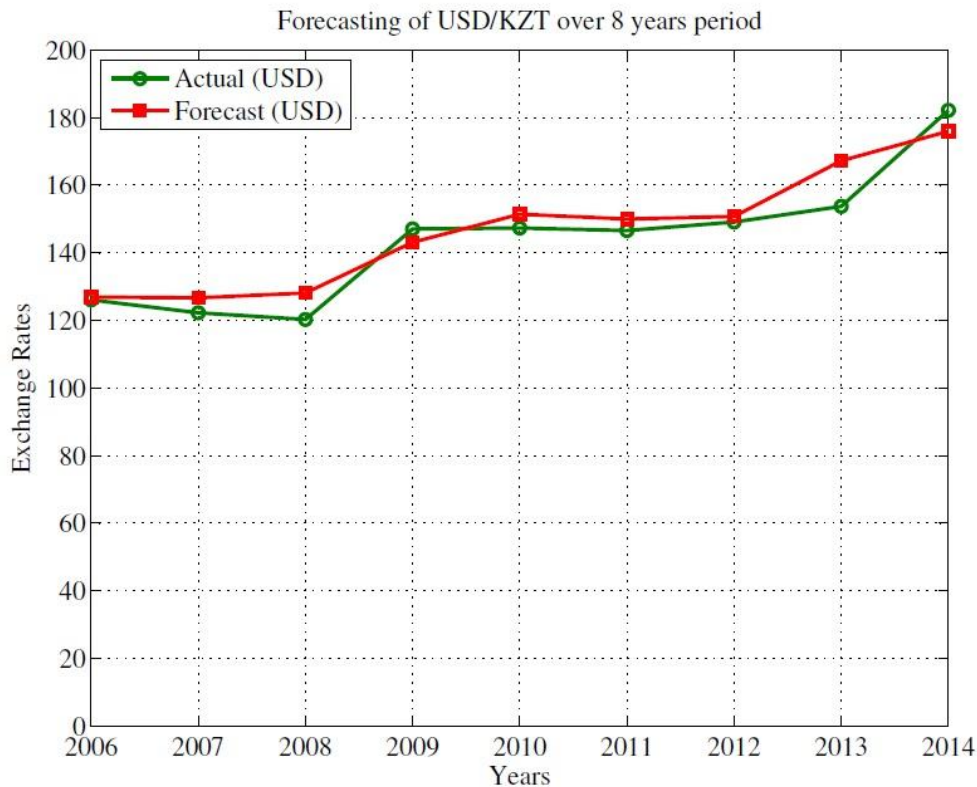

**Figure 2. Actual and forecast values of the USD/KZT exchange rates**

Figure 3 illustrates the actual and the forecast values for EUR/KZT exchange rates for the same period as it was performed for USD/KZT. One can identify that the exchange rates rose significantly from 2008 to 2009 and from 2012 to 2013. These significant changes were more than 16 per cent during the first increase, and more than 5 per cent during the second. It can be clearly seen that the developed forecast model is relatively close to the actual data, even during the drastic changes. Finally, the comparison for SGD/KZT exchange rates is shown in Figure 4.

As can be seen, developed ARIMA time series model for SGD/KZT was able to forecast the exchange rate values more accurately than the previous forecasts for other two currencies. To sum up, the actual and the developed ARIMA forecast values of exchange rates for three currencies with respect to Kazakh tenge were demonstrated and compared.

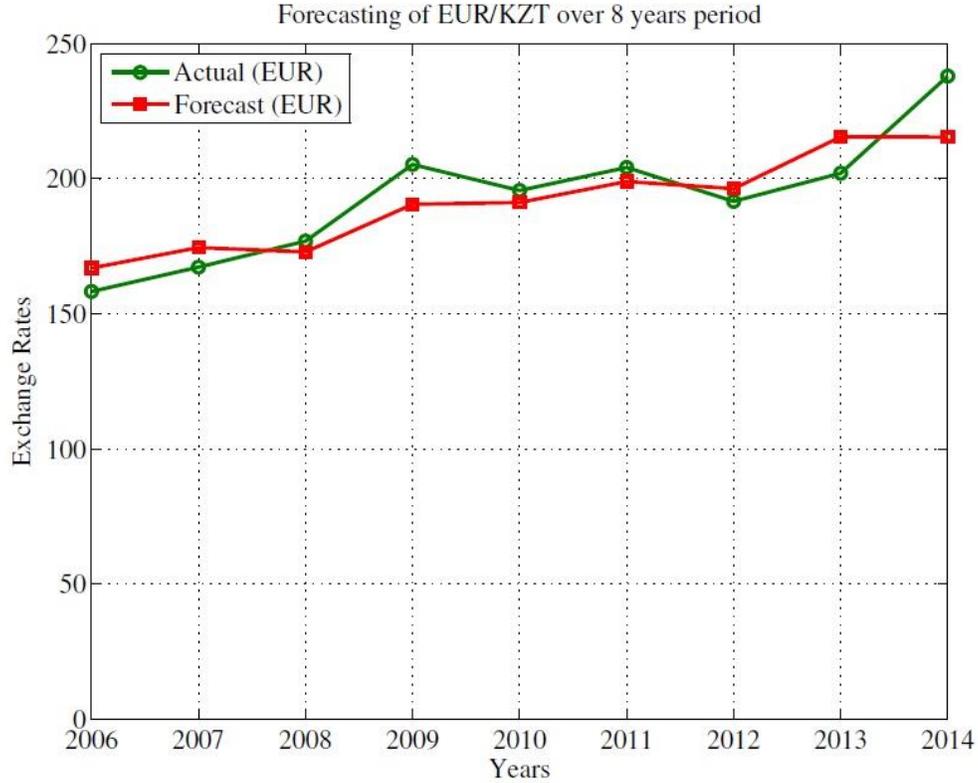

**Figure 3. Actual and forecast values of the EUR/KZT exchange rates**

Nowadays, one can evaluate forecast performance by applying it to non-statistical methods such as MAPE, RMSE and MAE [27]. These performance metrics were calculated as following [28]:

$$MAE = \frac{1}{k}\sum_{t=1}^{k}|\hat{X}_t - X_t|, \qquad (2)$$

$$MAPE = \frac{100}{k}\sum_{t=1}^{k}\left|\frac{\hat{X}_t - X_t}{X_t}\right|, \qquad (3)$$

$$RMSE = \sqrt{\frac{1}{k}\sum_{t=1}^{k}(\hat{X}_t - X_t)^2}, \qquad (4)$$

where k is the number of values, $X_t$ is actual exchange rate, $\widehat{X_t}$ is forecast exchange rate, and t is time. After calculation of the above mentioned performance metrics, the evaluation of the forecast results developed by time series ARIMA model are presented in Table 1. As can be seen, all values of the calculated performance metrics are relatively close to the actual data, with minimum error as 3.69 for Singapore dollar and maximum of 11.28 for euro. The error values for euro were more than the values for Singapore dollar. This can be explained by the fact that exchange rates for euro were fluctuating more than the Singapore dollar from 2006 to 2014, which can be observed from Figure 1. As for the adequacy of the performance metrics, the evaluation of ARIMA model via MAPE shows the minimum error values for all three currencies. For example, the MAPE was 4.45 per cent for US dollar, 4.64 per cent for euro, and 3.69 per cent for Singapore dollar. These results are in a good correlation with the results reported previously by Ahmed, S. et al. [28], where the MAPE values were within the range from 3.34 to 4.34 per cent. In addition, the MAPE values estimated in current study seem to be close to the results reported by Omane-Adjepong et al. [29], where the seasonal ARIMA models where implemented to forecast short term inflation. To sum up, the ARIMA model based forecasts of exchange rates for US dollar, euro and Singapore dollar with respect to Kazakh tenge were evaluated and showed relatively adequate results in comparison with the actual data.

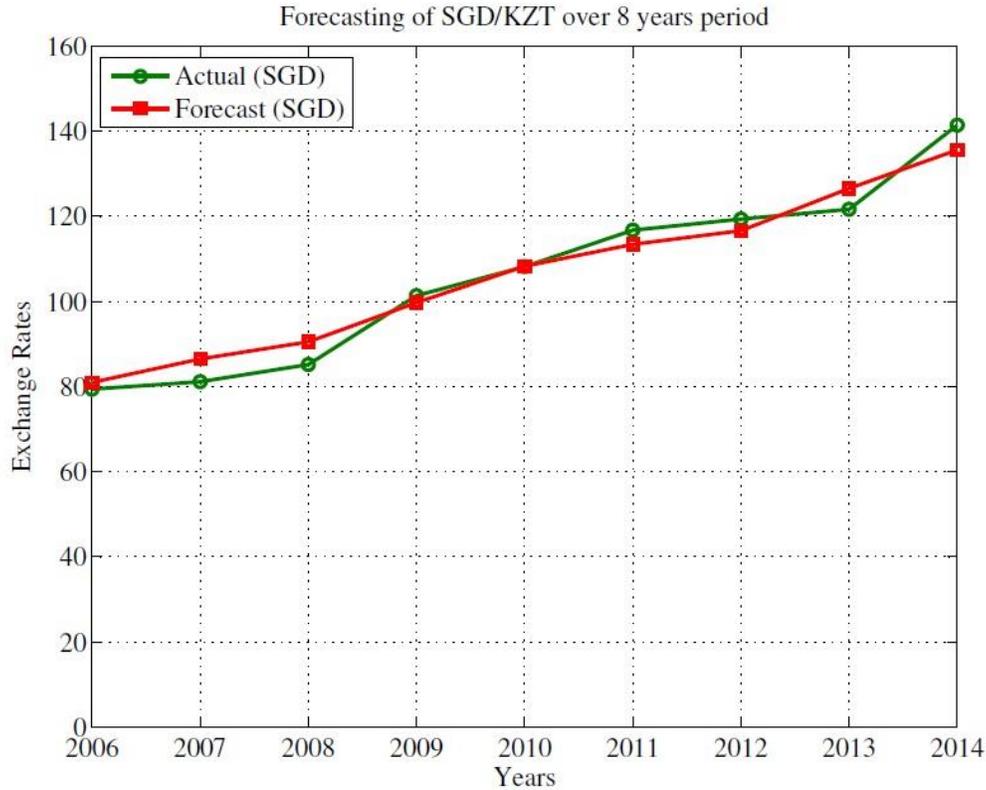

**Figure 4. Actual and forecast values of the SGD/KZT exchange rates**

| Currency | Performance Metrics | | |
|---|---|---|---|
| | MAE | MAPE | RMSE |
| **USD** | 6.540478 | 4.445373 | 7.912838 |
| **EUR** | 9.289714 | 4.63757 | 11.27605 |
| **SGD** | 3.692546 | 3.689833 | 4.202989 |

**Table 1. Performance metrics for USD/KZT, EUR/KZT, SGD/KZT**

## CONCLUSION

ARIMA technique for forecasting currency exchange rates of Kazakh tenge against three other currencies such as US Dollar (USD), Euro (EUR), and Singapore Dollar (SGD) was applied over the period from 2006 to 2014. The MATLAB software was utilized for prediction of the exchange rates. The ARIMA technique was presented and three main steps for constructing the model were identified, namely, Identification, Estimation, and Model checking. Furthermore, the forecast model was estimated and compared with the actual data for all three currencies. The effectiveness of the forecast model results was compared with mean absolute error (MAE), mean absolute percentage error (MAPE) and root mean squared error (RMSE). Results showed that the MAPE values for all three currencies were the smallest, i.e. the most effective. The MAPE values calculated in current work share certain similarities with the previously reported studies, which were compared as well.